\documentclass[printnumbers,amsmath,amssymb,twocolumn]{revtex4-2}
\usepackage{graphicx}
\usepackage{color}
\usepackage{float}
\usepackage{epstopdf}
\usepackage{hyperref}
\usepackage{booktabs}

\begin{document}

\title{Correlation between  the boson peak frequency and   transverse Ioffe-Regel limit  in four-dimensional structural glasses}

 \author{ Licun Fu$^{1}$, Xinyu Chang$^{1}$,   and Lijin Wang$^{1,*}$ \\
$^{1}${School of Physics, State Key Laboratory of Opto-Electronic Information Acquisition and Protection Technology, Anhui University, Hefei 230601, China}\\
$^{*}$ Corresponding author,  Email: lijin.wang@ahu.edu.cn\\
}

\begin{abstract}
The emergence  of excess vibrational modes over  the Debye prediction, typically manifested as the well-known  boson peak in the plot of  vibrational density of states scaled by the Debye prediction, has become  a hallmark of various amorphous solids.  The origin of the boson peak has been attracting considerable attention but is still under  debate. A popular view is  that  the position of the boson peak coincides well with that of the  Ioffe-Regel limit for transverse modes in  both two- and three-dimensional  glasses,  which is primarily derived from  simulation studies of model structural glasses.   However, it remains unknown whether the proposed coincidence  could be generalized to higher spatial dimensions, and addressing this  could contribute to the advancement of relevant  phenomenological theories.  Here,   we  find that  the transverse Ioffe-Regel  limit frequency  is higher than and  not proportional to the boson peak frequency in  our   studied four-dimensional glasses.  Our findings therefore suggest that the proposed coincidence between  the boson peak frequency and the transverse Ioffe-Regel limit  depends on spatial dimensions, which was not anticipated   previously. 

%It has been demonstrated that these excess modes play an important role in understanding  low-temperature thermal and mechanical properties of glasses. 

\end{abstract}

\date{\today}

\maketitle

\section{Introduction}
Glasses have long played a pivotal role in  the development of human civilization.  Nevertheless, the  nature of glasses continues to represent one of the most challenging scientific questions~\cite{Berthier_2011, Wang-ropp,lerner-JCP-review}.  Compared to crystalline materials, glasses exhibit distinct  low-temperature thermodynamic properties~\cite{experim_Zeller1971,thoery_Anderson1972,review_thermal2002}, particularly in terms of heat capacity and thermal conductivity.   One promising approach to understand these thermodynamic properties is to  decipher   low-frequency vibrational  peculiarities of glasses~\cite{Wang-ropp,lerner-JCP-review,Flenner_SM2020,NX_enrgytrans2009,ramos_book}. 

For crystals, the vibrational density of states (VDOS) at low frequencies $\omega$ follows the Debye prediction as $D(\omega)=A_{\rm D}\omega^{d-1}$ in $d$ dimensions with $A_{\rm D}$ the Debye level~\cite{Kittel}. However, the low-frequency VDOS in various glasses has been consistently observed to exceed the Debye prediction~\cite{Wang-ropp,lerner-JCP-review,mizuno_bigsystem_dos_pnas,wlj_dos_nc2019,Wang2021prl,Xu2010EPL,lerner_prl2016,wang_3d_jcp2022,wang_2d_jcp2023,Wang_prl2014}, typically manifested as an apparent peak in the plot of $D(\omega)/\omega^{d-1}$ versus $\omega$. This peak is commonly referred to as the boson peak, and the corresponding frequency is known as the boson peak frequency  $\omega_{\rm BP}$. Importantly, the temperature scale  corresponding to $\omega_{\rm BP}$  was found to be approximately located within the plateau region of thermal conductivity's temperature dependence  as well as within the bump  in the plot of heat capacity scaled by temperature squared~\cite{ramos_book}.  
Therefore, elucidating the nature of vibrational modes in the vicinity of the boson peak may provide valuable insights into the understanding of thermodynamic anomalies. Indeed, this objective represents one of the primary motivations behind numerous studies attempting to elucidate the origin of the boson peak or its correlation with other  characteristics~\cite{Wang-ropp}.  Over the decades, a variety of explanations have been proposed; however, no consensus has yet been reached regarding the origin of the boson peak~\cite{BPorigin-Zaccone-PRL, BPorigin-Schirmaher-PSSB, BPorigin_yuanchao-NP,BPorigin-Rufﬂe-PRL-TwoModels,BPorigin-Schober-PRL,Schrimacher_prl2007,Massimo_prl2021,Nie-Frontier,Jiang-NP}.

Due to the inherent  disorder and anharmonicity in glasses, phonons—even at extremely  low frequencies—experience  damping when propagating through glasses, as evidenced by a decreasing mean free path or lifetime of phonons with increasing  frequencies~\cite{Monaco-pnas-2009,ikeda_pre2018,Lemaitre-NM,Wang2019SMattenuation,Wang2020softmater,lerner2019JCP, Fupre, Fucpb,Szamel-jcp-2022,Szamel-SA,Szamel-jcp-2025}. When the mean free path of phonons decreases to  half their wavelength, the  frequency $\omega_{\rm IR}$  corresponding to the  Ioffe-Regel limit   is reached~\cite{silica-model-1,IR}.  Therefore, there will be no well-defined phonons any longer at frequencies  above $\omega_{\rm IR}$. Moreover, $\omega_{\rm IR}$ has been claimed to coincide with the crossover from weak to strong scattering of phonons in glasses~\cite{random-matrix-prb,random-lattice-prb,Monaco-pnas-2009,silica-model-1}.

Interestingly,  over the past two decades, one popular view   is that  $\omega_{\rm BP}$ corresponds to  the  transverse Ioffe-Regel  limit  $\omega_{\rm IR,T}$, which was mainly supported by  simulation studies  of two- (2D)  and three-dimensional (3D)  structural glasses~\cite{bp_shintani_NM2008,Monaco-pnas-2009,Mizuno-pnas-2014,Mizuno-pre-2018,Beltukov-bond-pre-2016,Schirmacher-SciRep,BPorigin_yuanchao-NP,MG-PRB-2012,Mizuno-gel-2022}. 
There are  several  additional simulation studies observing that $\omega_{\rm BP}$ coincides with both  transverse and longitudinal ($\omega_{\rm IR,L}$) Ioffe-Regel  limits in  model vitreous silica  glasses~\cite{silica-model-1,silica-model-2} and model gels~\cite{Mizuno-gel-2022}, where the equality $\omega_{\rm IR,T}=\omega_{\rm IR,L}$ was followed.  Note that the  inequality   $\omega_{\rm IR,T}<\omega_{\rm IR,L}$ has been observed in most model structural  glasses~\cite{bp_shintani_NM2008,xipeng-PRL}.  However,  simulation studies~\cite{Jiangzhehua-review,Nie-Frontier} of mass-spring networks where specific types of disorder could be tuned artificially, suggest that it's  possible to observe the  inequality between $\omega_{\rm IR,T}$ and  $\omega_{\rm BP}$ under appropriate conditions, whereas no such inequality has ever been observed in model structural glasses where multiple types of disorder coexist.  Additionally, there appears to be no consensus on the relation between the Ioffe-Regel limit and the boson peak based on  experimental observations~\cite{experiment-PRL-Monaco-IReqBP, experiment-PRL-Ruocco-comment,experiment-PRL-Ruocco-IRgtBP,experiment-PRL-Rufﬂe-comment,experiment-JCP-Ruta-IRgtBP,experiment-PRB-Ramos-IRgtBP,experiment-PRL-Baldi-IReqBP}.

To the best of our knowledge, nearly all numerical investigations of the correlation between the transverse Ioffe-Regel limit and  the boson peak are  done in 2D and/or   3D model structural  glasses. And these  studies consistently found $\omega_{\rm IR,T}=\omega_{\rm BP}$. Then, it is natural to consider what might happen to their relation when moving into higher spatial dimensions, and addressing this is  important to the development of relevant phenomenological theories. 

In this work, we performed computer simulation studies in  four-dimensional (4D) structural glasses.   Unexpectedly,  we find the equality  $\omega_{\rm IR,T}=\omega_{\rm BP}$ does not hold any more in 4D glasses studied. Specifically, we  observe that  $\omega_{\rm IR,T} > \omega_{\rm BP}$,  and $\omega_{\rm IR,T}$ is not proportional to $\omega_{\rm BP}$ in our studied 4D glasses.

\section{Simulation details}

We simulated  model glass formers with  two different types of interaction potentials:  (I) The inverse power law potential  and (II) the spring-like potential~\cite{Wang-sm-2012}. Each 4D  model glass former is  composed of a 50:50 binary mixture of $A$ particles and $B$ particles with equal mass (set to one in this work). Specifically, the interaction potential models   employed in our 4D simulations  are as follows: \\
(1) The  inverse power law (IPL) potential model.  In this model, the interaction between two particles $i$ and $j$  is    \textbf{$V(r_{ij})=  \big[(\frac{\sigma_{ij}}{r_{ij}}    )  ^{n}   +c_{0}   +c_{2}   (\frac{r_{ij}}{\sigma_{ij}}    )  ^{2}   +   c_{4}   (\frac{r_{ij}}{\sigma_{ij}}    )  ^{4} \Big ] G(r_{ij} ^{c}-r_{ij} )  $}. Here,   $G(r)$ is the Heaviside step function, $r_{ij}$ is the particle separation, $\sigma_{AA}=1.0$,  $\sigma_{BB}=1.4$,  and $\sigma_{AB}=1.18$.  The polynomial terms are  introduced to smooth  $V(r_{ij})$
up to its second  derivative at  the interaction cutoff  $r_{ij}^{c}=1.48 \sigma_{ij}$.  Here, we employed  $n=10$ (hereafter referred to as IPL-n10), and $n=14$ (IPL-n14).  \\
(2) The spring-like potential model. In this model, $V(r_{ij}) =\frac{1}{\alpha} (1-  \frac{\sigma_{ij}}{r_{ij}}    ) ^{\alpha}       G(\sigma_{ij}-r_{ij} ) $, where $\sigma_{AA}=1.0$, $\sigma_{BB}=1.4$,  and $\sigma_{AB}=1.2$. Here, we employed $\alpha = 2$ corresponding to the harmonic potential (hereafter referred to as HARM), and $\alpha = 2.5$ corresponding to the Hertzian potential (HERTZ).

\begin{figure}[t]
\includegraphics[width=0.5\textwidth]{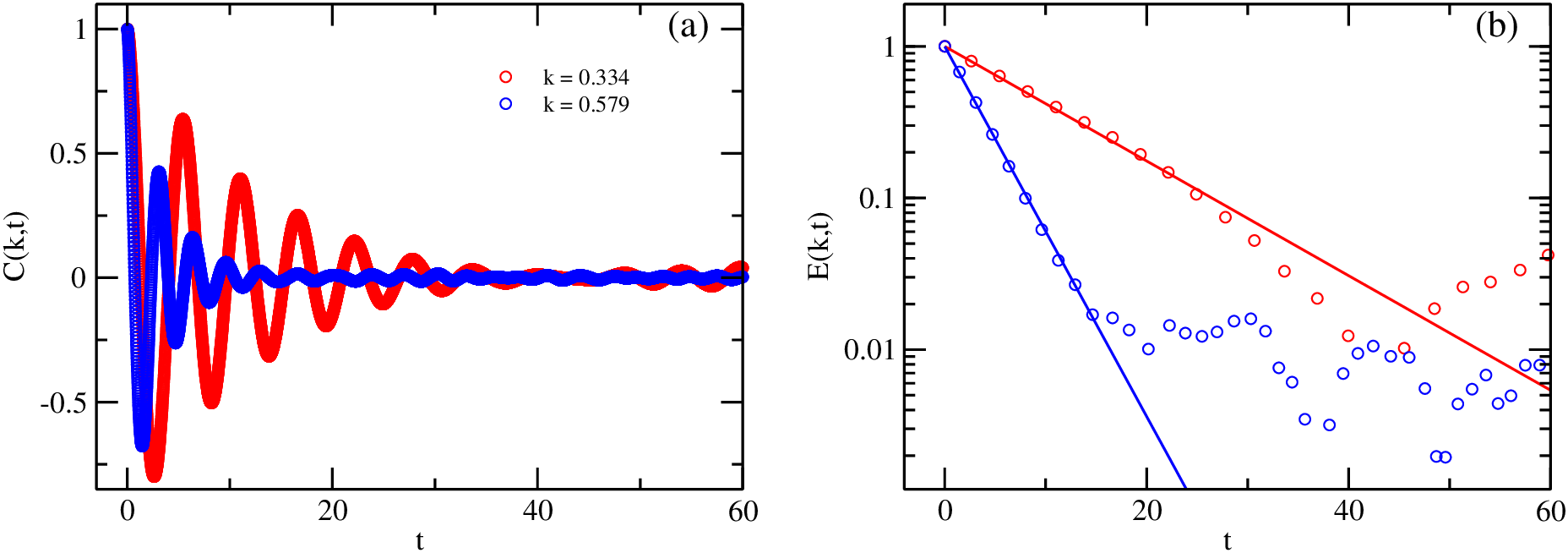}% Here is how to import EPS art
\caption{\label{fig:envelope} (a) Illustration of velocity correlation functions $C(k,t)$ at  wavevectors $k=0.334$ and $k=0.579$ in 4D glasses. (b) Time dependence of the envelope $E(k,t)$  determined by the peaks in the absolute value of $C(k,t)$ shown in (a).  
Data in (a) and (b) are obtained at $\rho=0.8$ in 4D IPL-12 glasses.  }
\end{figure}

 Different 4D model glass formers  were created  by varying the types of interaction potentials  and number densities $\rho=N/L^d$ with $L$ the side length and $N$ the total number of particles.  In particular,  the system sizes in 4D range from $20000$ to $300000$, depending on  number densities and interaction potential models. 
 %The number density for each interaction potential model is presented  in Table.~\ref{table}.

 To avoid redundancy and potential confusion, we omit here the simulation details for the 2D and 3D data presented in this work, which were obtained from IPL glass models with $n=12$ (hereafter referred to as IPL-n12).  For further details on 2D IPL glass models, readers are referred to Ref.~\cite{Wang2021prl}; for 3D IPL glass models, see Ref.~\cite{wlj_dos_nc2019}.

Zero-temperature ($T=0$) glasses  were  produced by quenching instantaneously   high-temperature liquids to $T=0$  via energy minimization  (here we use the fast inertial relaxation engine~\cite{fire}). The frequencies of vibrational modes were determined through the diagonalization (using math kernel library~\cite{mkl}) of the dynamic matrix derived from $T=0$ glasses. The vibrational density of states is given by $D(\omega)=\frac{    1 }{dN-d} \sum^{dN-d} _{l=1} \delta(\omega - \omega_{l})$, where $ \omega_{l}$ is the frequency corresponds to the $l$-th vibrational mode.   
\begin{figure}[t]
\includegraphics[width=0.48\textwidth]{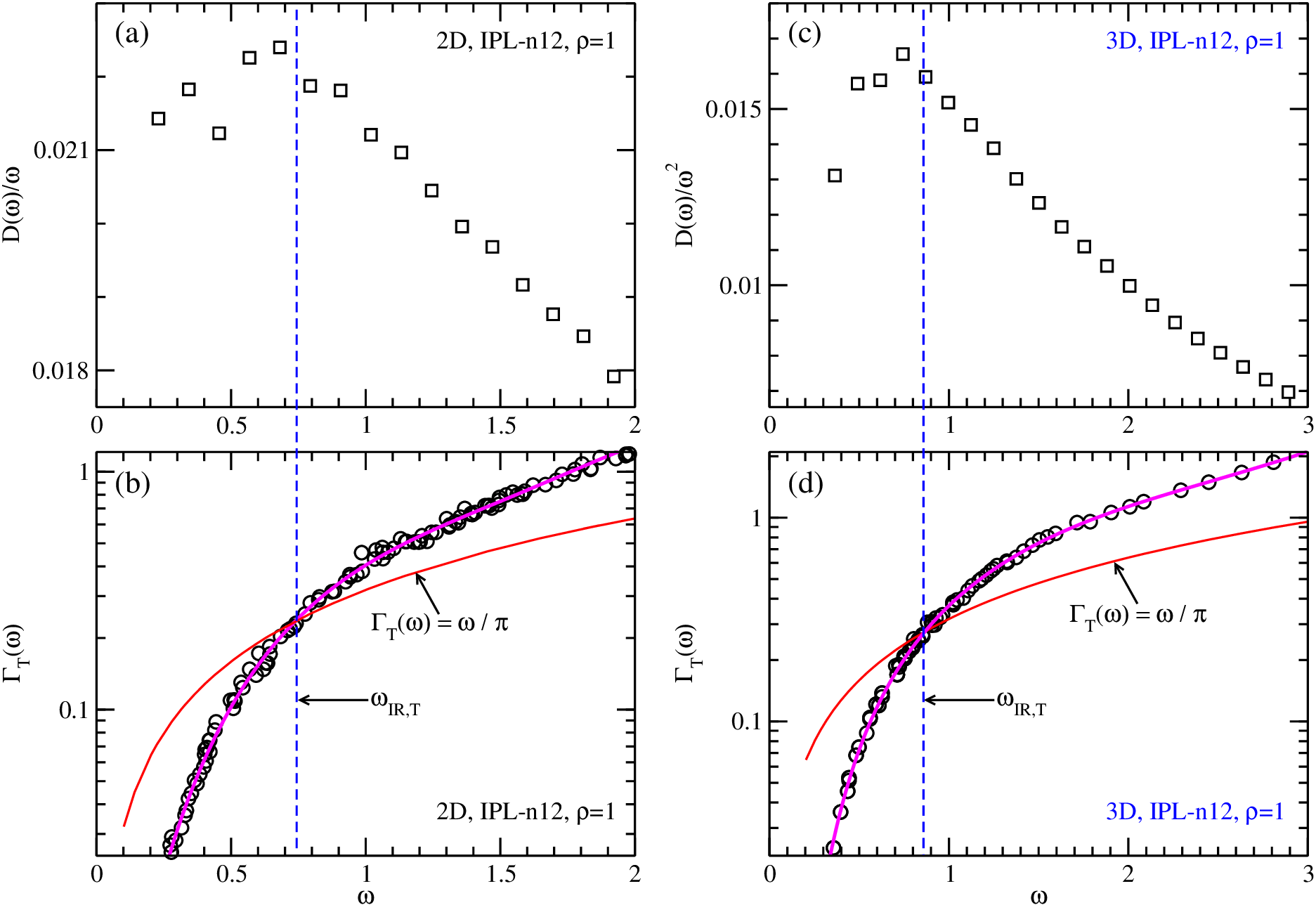}% Here is how to import EPS art
\caption{\label{fig:2D3D}   Reduced VDOS $D(\omega)/\omega$ in (a) and transverse sound attenuation  coefficient  $\Gamma_{\rm T} (\omega)$ in (b)  at $\rho=1$ in 2D IPL-12 glasses with $N$ ranging from 20000 to 50000. 
 Reduced VDOS $D(\omega)/\omega^2$ in (c) and transverse sound attenuation  coefficient  $\Gamma_{\rm T} (\omega)$ in (d)  at $\rho=1$ in 3D IPL-12 glasses with $N$ ranging from 48000 to 192000. The lines passing through  data points of $\Gamma_{\rm T} (\omega)$ in (b) and (d) are obtained from polynomial fits.    }
\end{figure}

 The transverse sound attenuation coefficient (or inverse lifetime)  $\Gamma_{\rm T} (\omega)$ was determined as follows.  First, the  initial velocity of  particle $i$ was set  to  $ \mathbf{ \dot{u   } }_{i} (t=0)     =   \mathbf{ b}  \sin(\mathbf{ k \cdot r}_{i} )$. Here,    $  \mathbf{b\cdot k }=0                    $,  $\mathbf{ k }$ is the wavevector,  $\mathbf{ b}$ is a unit vector,  and $\mathbf{ r }_{i}$  is the position vector of particle $i$ at $t=0$ in $T=0$ glasses.  Second,  we performed molecular dynamics (MD) simulations by solving the equation of motion~\cite{Lemaitre-NM} which  reads

\begin{equation}
      \mathbf{ \ddot{u}}_{i}(t) =  \mathbf{ \dot{u} }_{i}(t=0) \delta(t)   -  \sum\limits^N \limits_{j=1}D_{ij} \mathbf{ u}_{j}(t),  
      \label{6}
\end{equation} 
where $D_{ij}$ is the dynamic matrix of $T=0$ glasses, and    $\mathbf{ u }_{i}(t)$  represents the displacement of particle $i$  at  $t$ relative to its initial position.  During MD simulations, we  calculated the velocity correlation function 
\begin{equation}
 C (k,t) =\Bigg\langle
\frac{       \sum\limits^N \limits_{i=1}                 \mathbf   {\dot{ u }} _{i}(0)  \cdot              \mathbf   {\dot{ u }} _{i}(t)  }
{    \sum\limits^N \limits_{i=1}                 \mathbf   {\dot{ u }} _{i}(0)  \cdot              \mathbf   {\dot{ u }} _{i}(0)}\Bigg\rangle. \label{7}
\end{equation} 
The transverse sound attenuation $\Gamma_{\rm T}(\omega)$ and the corresponding frequency $\omega$    could be  obtained by  fitting the  $C(k,t)$ data to  
\begin{equation}
 C (k,t) =\exp(-\Gamma_{\rm T}(\omega) t/2)\cos(\omega t).
\label{eq1}
\end{equation}
Note that  $\omega$ depends on $k$, and so does  $\Gamma_{\rm T}(\omega)$. $k$  in this work  ranges approximately from 0.13 to 1.58, and the specific values of $k$ (satisfying periodic boundary conditions) depend on systems examined.

\section{Results}

 \begin{figure}[t]
\includegraphics[width=0.49\textwidth]{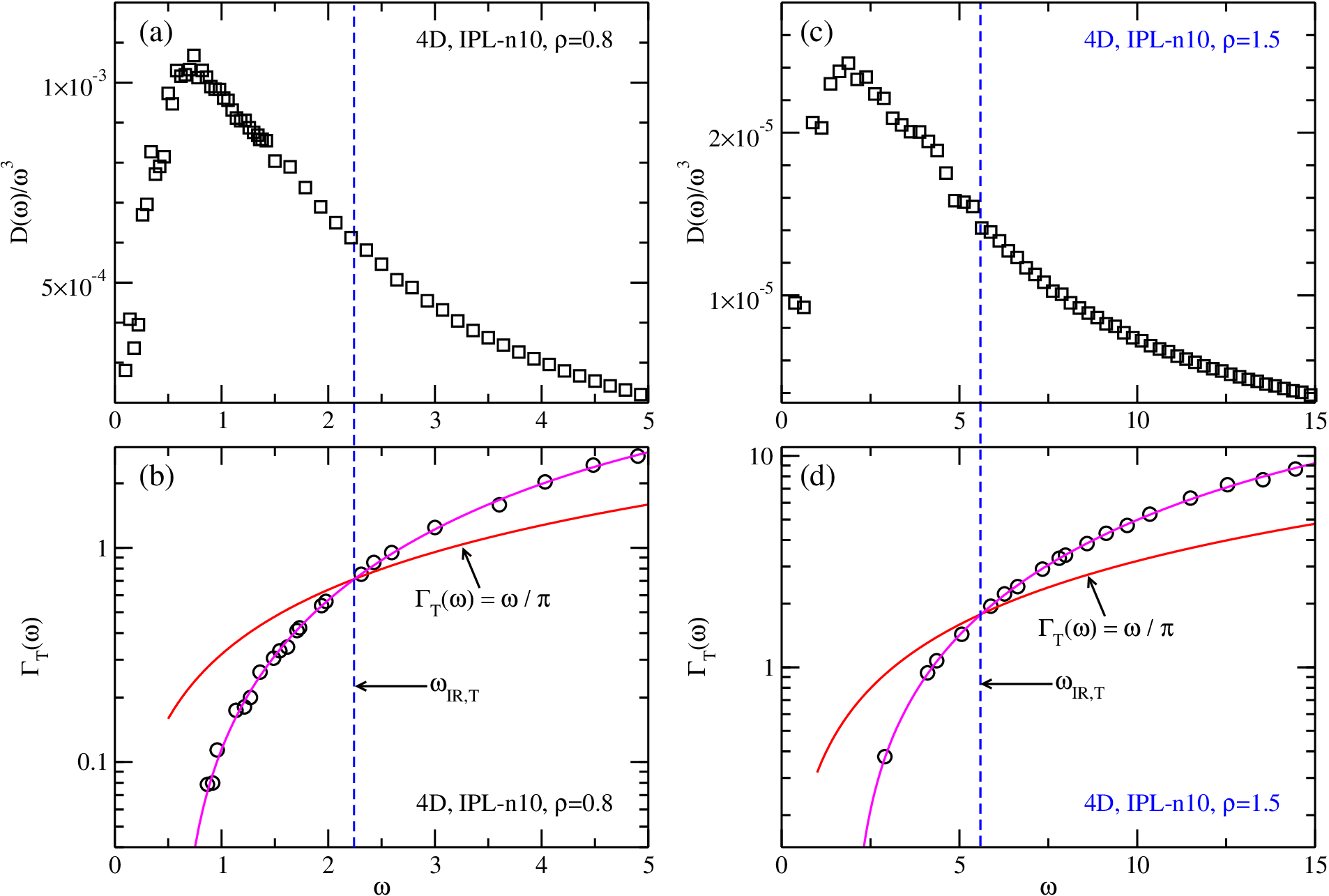}% Here is how to import EPS art
\caption{\label{fig:4Dn10}  Reduced VDOS $D(\omega)/\omega^3$ in (a) and (c) and transverse sound attenuation  coefficient  $\Gamma_{\rm T} (\omega)$ in (b) and (d)  in 4D IPL-10 glasses. 
Data in (a) and (b) are obtained at $\rho=0.8$, and (c) and (d) are at $\rho=1.5$.
The lines passing through  data points of $\Gamma_{\rm T} (\omega)$ in (b) and (d) are obtained from polynomial fits.  }
\end{figure}

Figure~\ref{fig:envelope}(a)  illustrates the time evolution of $C (k,t)$ calculated at two different $k$ in 4D glasses. One could observe that  $C(k,t)$ decays faster at  a higher $k$.  In addition,  the  decay at long time is abnormal, which could be seen more clearly from the time evolution of the envelopes of $C(k,t)$, $E(k,t)$,  see Fig.~\ref{fig:envelope}(b).  Here, $E(k,t)$   corresponds to all   peaks in $|C(k,t)|$ and exhibits significant fluctuations at long time scales at each fixed $k$.  Note that such anomalous decay of $E(k,t)$ at long time has  been observed as well in both 2D and 3D glass systems, which was  demonstrated to be due to finite-size effects~\cite{Wang2019SMattenuation,Lemaitre-NM,Fucpb}.   

To eliminate the finite-size effects,   a restricted envelope fit method has been introduced in recent years to get $\Gamma_{\rm T}(\omega)$ in simulations in  2D and 3D glasses~\cite{Wang2019SMattenuation,Wang2020softmater,Fupre,Fucpb}.  In this work, we employed the same method to get  $\Gamma_{\rm T}(\omega)$ in 4D glasses.  Specifically, as illustrated in  Fig.~\ref{fig:envelope}(b), the appropriate short-time $E(k,t)$ data are selected to fit to the equation  $E(k,t)=\exp(-\Gamma(\omega) t/2)$.  We observe that  the  resulting  $\Gamma_{\rm T}(\omega)$   exhibits no   finite-size effects in 4D glasses.   Unless otherwise specified, all $\Gamma_{\rm T}(\omega)$ data presented below are derived from measurements across different system sizes.  Additionally, the frequency $\omega_{\rm IR,T}$  corresponding to the transverse Ioffe-Regel limit is determined as the value when $\pi\Gamma_{\rm T}(\omega)=\omega$~\cite{bp_shintani_NM2008}.    

The investigation of the correlation between the Ioffe-Regel limit and the boson peak has attracted considerable attention. Based on  experimental studies~\cite{experiment-PRL-Monaco-IReqBP, experiment-PRL-Ruocco-comment,experiment-PRL-Ruocco-IRgtBP,experiment-PRL-Rufﬂe-comment,experiment-JCP-Ruta-IRgtBP,experiment-PRB-Ramos-IRgtBP}, their correlation remains an unresolved issue that requires further investigations.   However, a growing consensus has emerged  from simulation studies over the past two decades~\cite{bp_shintani_NM2008,Monaco-pnas-2009,Mizuno-pnas-2014,Mizuno-pre-2018,Beltukov-bond-pre-2016,Schirmacher-SciRep,BPorigin_yuanchao-NP,MG-PRB-2012,Mizuno-gel-2022,silica-model-1,silica-model-2}.
These simulation studies   suggested  that the transverse Ioffe-Regel limit coincides with the boson peak frequency, i.e., $\omega_{\rm IR,T}=\omega_{\rm BP}$, irrespective of  interaction potentials or spatial dimensions (at least in 2D and 3D).    Regarding  the relation between $\omega_{\rm IR,T} $ and $\omega_{\rm IR,L}$,  it was found  $\omega_{\rm IR,T}< \omega_{\rm IR,L}$ in most model structural glasses~\cite{bp_shintani_NM2008}, whereas simulation studies on model silica glasses~\cite{silica-model-1,silica-model-2}  reported  $\omega_{\rm IR,T}=\omega_{\rm IR,L}=\omega_{\rm BP}$, suggesting that both the transverse and longitudinal  Ioffe-Regel limits coincide with the boson peak.   Despite possible variations in the relative positions of $\omega_{\rm IR,T} $ and $\omega_{\rm IR,L}$ across different glasses, $\omega_{\rm IR,T}=\omega_{\rm BP}$ appears to hold consistently in 2D and 3D glasses. 

\begin{figure}[t]
\includegraphics[width=0.48\textwidth]{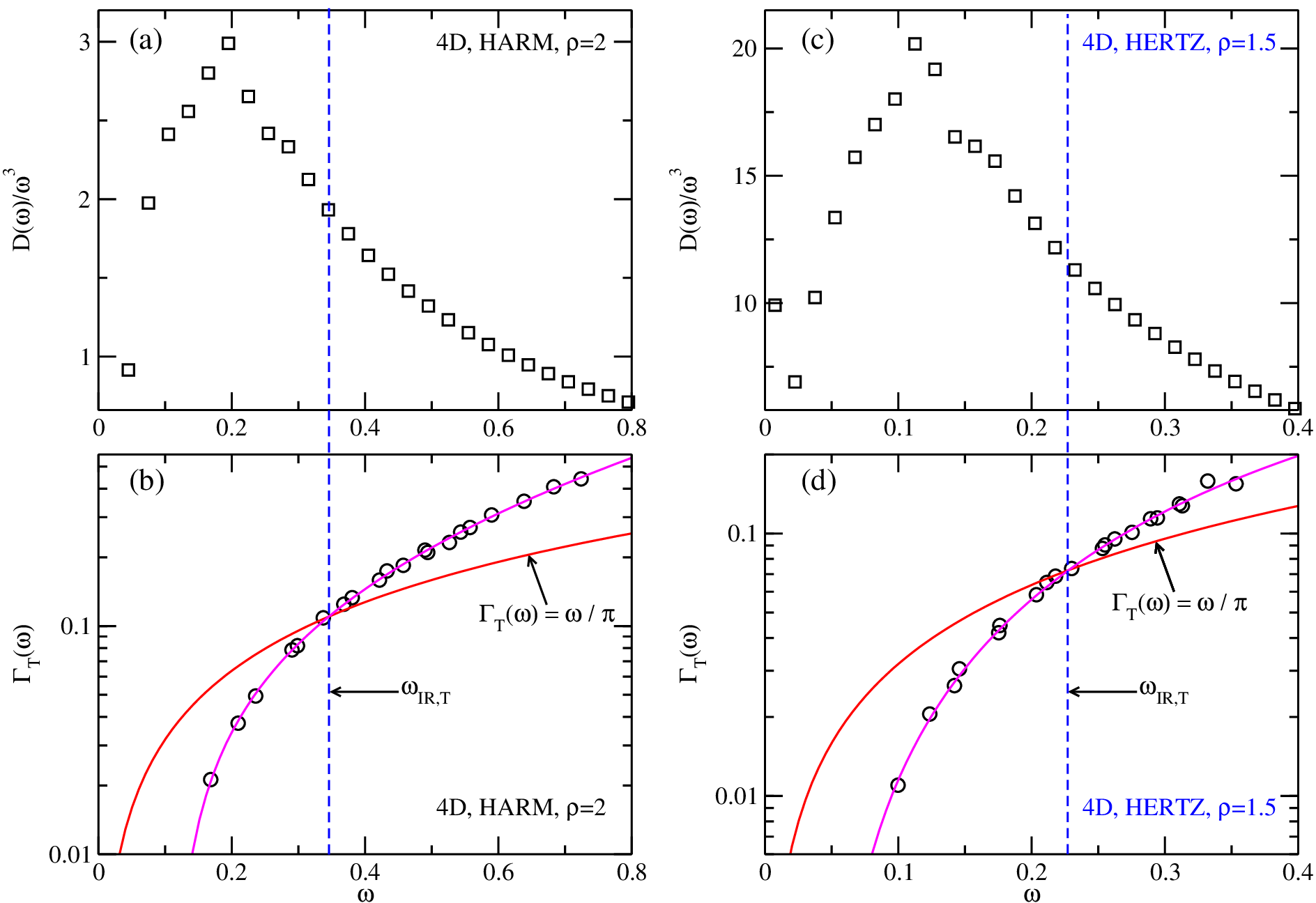}% Here is how to import EPS art
\caption{\label{fig:4DHARM} Reduced VDOS $D(\omega)/\omega^3$ in (a)   and (c) and transverse sound attenuation  coefficient  $\Gamma_{\rm T} (\omega)$ in (b) and (d)  in 4D glasses with spring-like potentials. Data in (a) and (b) are calculated at $\rho=2$ in HARM glasses, and (c) and (d) are at $\rho=1.5$ in HERTZ glasses. The lines passing through data points of $\Gamma_{\rm T} (\omega)$ in  (b) and (d) are obtained from polynomial fits.  }
\end{figure}

First, we revisit the correlation between  $\omega_{\rm IR,T}$ and $\omega_{\rm BP}$  in our studied 2D IPL-12 model glass formers in  Figs.~\ref{fig:2D3D}(a) and (b)  and in  3D IPL-12 ones  in  Figs.~\ref{fig:2D3D}(c) and (d).  Note that Figs.~\ref{fig:2D3D},  ~\ref{fig:4Dn10}, and ~\ref{fig:4DHARM} are  presented using a consistent plotting format for clarity.   $\omega_{\rm BP}$ is typically identified as the peak position in the reduced plot of  $D(\omega)/\omega^{d-1}$ versus $\omega$, as shown in Figs.~\ref{fig:2D3D} (a) and (c).  $\omega_{\rm IR, T}$ is determined by the intersection between the line representing $\Gamma_{\rm T}(\omega)=\omega/\pi$ and the one representing the polynomial fit to the  calculated $\Gamma_{\rm T}(\omega)$ data, see Figs.~\ref{fig:2D3D} (b) and (d).  
One could observe that the vertical dashed line (representing $\omega_{\rm IR, T}$) passing through  Fig.~\ref{fig:2D3D}(a) and  (b)   is  located near the boson peak position ($\omega_{\rm BP}$) in Fig.~\ref{fig:2D3D}(a), thus suggesting $\omega_{\rm IR,T} \approx \omega_{\rm BP}$. Similar  results can be observed in  Figs.~\ref{fig:2D3D} (c) and  (d) for 3D glasses.   Therefore, the previously reported equality $\omega_{\rm IR,T} = \omega_{\rm BP}$ is further supported by our studied  2D and 3D model glasses.  

 \begin{figure}[t]
\includegraphics[width=0.45\textwidth]{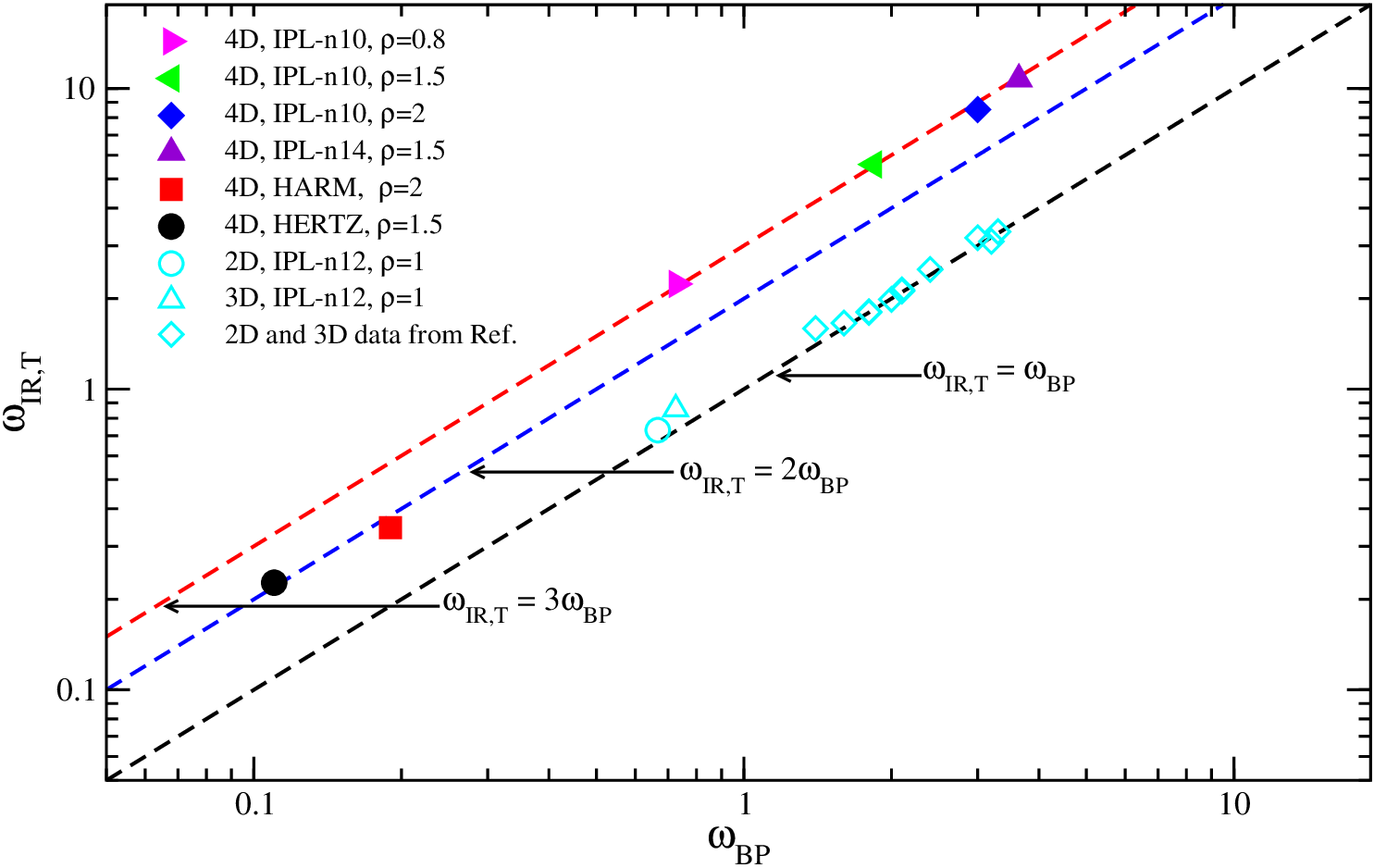}% Here is how to import EPS art
\caption{\label{fig:BPIR}  Correlation between the boson peak frequency $\omega_{\rm BP}$ and   the  transverse Ioffe-Regel limit frequency $\omega_{\rm IR,T}$.  Some 2D and 3D data shown here are from Ref.~\cite{bp_shintani_NM2008}. The values of $\omega_{\rm BP}$ and  $\omega_{\rm IR,T}$ for each 4D system shown here are listed in Table.~\ref{table} 
}
\end{figure}

A compilation of   simulation studies on 2D and 3D model structural glasses indicates that $\omega_{\rm IR,T} = \omega_{\rm BP}$.   Then, it is natural to examine whether the relation $\omega_{\rm IR,T} = \omega_{\rm BP}$  still holds in  higher spatial dimensional glasses.  To the best of our knowledge, the correlation between $\omega_{\rm IR,T}$ and  $\omega_{\rm BP}$ in  systems with $d>3$  has yet to be investigated, though there have been some studies of VDOS in such systems~\cite{Shimada_PRE2020, patrick,Wang-CPB-4D-dos,lerner_dos_234d_prl2018}.
Here, we performed computer simulations to  investigate their correlation  in 4D model structural glasses with  different types of interaction potentials and/or number densities. 

We are now in a position to demonstrate our main findings in this work.   We first show results in  4D IPL glasses.  We compare  $\omega_{\rm IR,T}$ and  $\omega_{\rm BP}$   in 4D IPL-10 systems in Fig.~\ref{fig:4Dn10}.    We show  results from  4D IPL-10 systems at $\rho=0.8$ in Figs.~\ref{fig:4Dn10} (a) and (b); unexpectedly,  one could  observe that the vertical line  representing  $\omega_{\rm IR,T}$ is far away from the position of the boson peak in Fig.~\ref{fig:4Dn10} (a).  This is very different from the observations from Fig.~\ref{fig:2D3D} for 2D and 3D  glasses.
The similar conclusion can be drawn  from  Figs.~\ref{fig:4Dn10} (c) and (d)  for 4D IPL-10 systems  at $\rho=1.5$.     These observations in 4D IPL systems thus suggest that $\omega_{\rm IR,T}$  appears at a  higher frequency than $\omega_{\rm BP}$, which is our main finding in this work.  

Additionally,  our  finding  of $\omega_{\rm IR,T} > \omega_{\rm BP}$ is further supported by  results from 4D glasses with spring-like potentials. The studied systems with such potentials are 4D HARM glasses at $\rho=2.0$ in Figs.~\ref{fig:4DHARM} (a) and (b) and 4D HERTZ glasses at $\rho=1.5$ in Figs.~\ref{fig:4DHARM} (c) and (d). Apparently, the plots in Fig.~\ref{fig:4DHARM} indicate that  $\omega_{\rm IR,T}$  is greater than  $\omega_{\rm BP}$.

%%%%%%%%%%%%%%%%%%%%%%%%%%%---------------
\begin{table}[t]  % 优先当前位置(h)，其次顶部(t)，然后底部(b)，最后浮动页(p)
  \centering
 \caption{The boson peak frequency $\omega_{\rm BP}$, the  transverse Ioffe-Regel limit frequency $\omega_{\rm IR,T}$, and the number density $\rho$ in 4D systems studied in this work. Here, $\omega_{\rm BP}$ is determined as the peak position in the plot of $D(\omega)/\omega^3$ versus $\omega$, whereas $\omega_{\rm IR,T}$ is the intersection between the line representing $\Gamma_{\rm T}(\omega)=\omega/\pi$ and the one representing the polynomial fit to the  $\Gamma_{\rm T}(\omega)$ data, as shown in  Figs.~\ref{fig:4Dn10} and ~\ref{fig:4DHARM}.   }
\label{table}

\begin{tabular}{c@{\hspace{35pt}}c@{\hspace{35pt}}c@{\hspace{35pt}}c}
\toprule
  \midrule
  System &  $\rho$ & $\omega_{\rm BP}$ &  $\omega_{\rm IR,T}$   \\  
  \midrule
  
 4D IPL-n10&  0.8  &  0.73 & 2.24   \\
 \addlinespace
 4D IPL-n10& 1.5 & 1.84  &  5.59 \\
 \addlinespace
 4D IPL-n10&  2.0   &  3.00  &  8.51  \\
  \addlinespace
4D IPL-n14& 1.5   &  3.64  & 10.74  \\
 \addlinespace   4D HARM&  2.0  &  0.19 &  0.35  \\
 \addlinespace
4D  HERTZ&  1.5   &  0.11 &  0.23   \\
 \bottomrule
\end{tabular}
 \end{table}
%%%%%%%%%%%%%%%%%%%%%%%%%%%%%%%%%%%%%

\section{Conclusion}

In this work,  we  investigate the relation between the boson peak and   the  transverse Ioffe-Regel limit in 4D glasses. Our main finding is that  the transverse Ioffe-Regel limit is reached at a much higher frequency than the boson peak frequency in all our studied 4D glasses.   Specifically, Fig.~\ref{fig:BPIR}  summarizes the relation between  $\omega_{\rm IR,T}$ and $\omega_{\rm BP}$ in  our studied 4D glasses as well as in   previously studied 2D and 3D glasses~\cite{bp_shintani_NM2008}. 
If  assuming $\omega_{\rm IR,T}=A\omega_{\rm BP}$,  one could observe that $A \approx1$ in  2D and 3D glasses.  However, in 4D systems, we find $A \approx 2 $ in HARM and HERTZ glasses, whereas $A \approx 3 $  in  IPL-10 and IPL-14  glasses. This indicates that  $\omega_{\rm IR,T} > \omega_{\rm BP}$, and  $\omega_{\rm IR,T}$ is not  proportional to $\omega_{\rm BP}$ as  the   proportionality coefficient $A$ is not constant. Note that the possibility that $A$ may assume other values including unity, in other 4D glasses not examined here cannot be ruled out.  Our current findings thus suggest that the previously proposed  coincidence between the boson peak and  the  transverse Ioffe-Regel limit  depends on spatial dimensions.

The   reported coincidence between the transverse Ioffe-Regel limit and the boson peak frequency in 2D and 3D model structural glasses is believed to offer a unique perspective for understanding the origin of the boson peak~\cite{bp_shintani_NM2008}.
However, it should be noted that a simulation study~\cite{Nie-Frontier} of  2D  mass-spring networks where specific types of disorder can be  tuned artificially, suggested that the relation $\omega_{\rm IR,T} = \omega_{\rm BP}$ is conditional, as it demonstrated the possibility of observing $\omega_{\rm IR,T} > \omega_{\rm BP}$.  We also notice that the heterogeneous elasticity  theory  predicts the possibility of observing  $\omega_{\rm IR,T} > \omega_{\rm BP}$  under appropriate conditions~\cite{Schrimacher_prl2007}.  However, such inequality has not been observed convincingly  in any simulation study of more realistic model structural glasses  where various types of disorder coexist.  Additionally, our observation  of $\omega_{\rm IR,T} > \omega_{\rm BP}$ in 4D glasses in this work   implies that it would be necessary to revisit whether $\omega_{\rm IR,T} > \omega_{\rm BP}$ can also be observed in specific 2D or  3D model structural glasses. 

%The spectrum of low-frequency excess vibrational modes (hybridizing weakly with phonon modes)   beyond the Debye prediction, has been observed to follow a universal scaling law across  2D, 3D, and 4D glasses~\cite{Wang-CPB-4D-dos,wang_3d_jcp2022,wang_2d_jcp2023}.   Moreover, low-frequency excess vibrational modes and sound attenuation have been shown to exhibit a quantitative correlation~\cite{Wang2019SMattenuation,Massimo_prl2021}.    

\section*{ACKNOWLEDGMENTS}

We acknowledge  the support from   National Natural Science Foundation of China (Grant Nos. 12522503 and 12374202), Anhui Projects (Grant Nos. 2022AH020009, S020218016 and Z010118169), and Hefei City (Grant No. Z020132009).  We  also acknowledge Hefei Advanced Computing Center, Beijing Super Cloud Computing Center, and the High-Performance Computing Platform of Anhui University for providing computing resources.

\section*{DATA AVAILABILITY}
The data that support the findings of this study are available from the corresponding authors upon reasonable request

\end{document}